# Sequential Labelling and DNABERT For Splice Site Prediction in *Homo Sapiens* DNA

**Muhammad Anwari Leksono[1], Ayu Purwarianti[1]**
[1]School of Electrical Engineering and Informatics, Bandung Institute of Technology, Bandung 40135, Indonesia

Corresponding author: Muhammad Anwari Leksono (e-mail: m_anwari@alumni.itb.ac.id, lm.anwari@gmail.com).

**ABSTRACT** Gene prediction on DNA has been conducted using various deep learning architectures to discover splice sites in order to discover intron and exon region. Recent predictions are carried out with models trained on sequence with splice site located in the middle of sequence. This case eliminates possibility of multiple splice sites existence in single sequence. This paper proposes sequential labelling model to predict splice sites regardless their position in sequence. Sequential labelling is carried out on DNA to determine intron and exon region and thus discover splice sites. Sequential labelling model named DNABERT-SL is developed on pretrained DNABERT-3. Both fine-tuning and feature-based approach are tested. DNABERT-SL is benchmarked against the latest sequential labelling model designed for mutation type and location prediction. While achieving F1 scores above 0.8 on validation data, both BiLSTM, BiGRU, and DNABERT-SL perform poorly on test data as indicated by their respective low F1 scores (0.498±0.184, 0.6±0.123, 0.532±0.245). Error and test results reveal that model experience overfitting and struggles to predict exon and splice site labels. Principal component analysis on token contextual representation produced by DNABERT-SL shows that the representation is not optimal to distinguish splice site tokens with non splice site tokens. Splice site motif observation conducted on test and training sequence shows that an arbitrary sequence with GT-AG motif can be both splice site in some sequences and normal nucleotides in other sequences. DNABERT-SL model cannot distinguish nucleotides acting as splice sites from normal ones.

## I. INTRODUCTION

Splice site prediction is a task in which DNA is analyzed to predict its splice site location to isolate exon from intron and therefore carry out protein prediction task. Numerous studies have been conducted to identify splice site in both classical machine learning and deep learning approaches. Classical methods such as Hidden Markov Model [1], Bayesian Networks [2], Random Forest [3], and SVM [4] have been used for splice site prediction. While classical methods provide acceptable result in their specific datasets, it is well known that classical methods require preprocessing to find optimal features [5]. Deep learning approach removes the need for manual feature selection since it by itself can learn the most optimal features for the task. By utilizing neural networks, such as CNN [6] and LSTM [7], splice site prediction task can be carried out with good results [8]–[13]. CNN has been popular choice for this problem and delivered apparent state-of-the-art performance and LSTM approach has been only introduced in recent works.

Recent works in splice site prediction make use of fixed-length sequence with splice site is located at center of sequence and has only one splice site. Various sequence lengths have been tested and it is inferred that flanking length is one of contributing factors to model performance aside from optimized CNN/RNN architecture. Left flanking region can be seen as information that determine whether next nucleotide is a splice site. The right flank contains certain information confirming whether the preceding character is splice site. In case of existing models, flanking region with certain length, such as $n$, is utilized under assumptions that there is no splice-site found before $n^{th}$ nucleotide on single gene. Furthermore, short sequences such as microRNA consists of only exons. Existing models can predict splice site inexistence correctly in the sequence. However, these models are trained to predict whether the sequence contains only exons. To improve model capabilities to recognize all-intron and all-exon sequence and splice site locations, sequential labelling model is proposed.

Our work proposes hypothesis that splice site prediction can be carried out by labelling each nucleotide in sequential manner. This labelling task is called sequential labelling. Sequential labelling has been utilized to solve natural language processing (NLP) problems such as NER, POS Tagging, or sentence chunking [14]. Latest development in NLP results in development of word embeddings and language models. Word embedding such as Word2Vec [15] and GloVe [16] have been popular choices to create sentence embedding to solve NLP problems. Language model is later developed to provide contextualized embeddings, with which polysemy word can be distinguished properly [17]. Recent developments of Attention [18], Transformers [19], BERT [20], which have delivered state-of-the-art language model, improve the performance of sequential labelling model.

Like natural language, genetic data can be regarded as sentences with different letters such as nucleotides or amino acids, biophysical, and biochemical rules. This similarity allows NLP methods such as embeddings and language models to be utilized in DNA analysis [21]–[24]. Embedding and language model are expected to fasten model training without compromising resulting model performance on multiple Bioinformatics problem.

## II. MATERIAL AND METHODS

### A. Data

Proposed sequential labelling model uses pretrained DNABERT-3 [23] and NCBI RefSeq data [25]. NCBI RefSeq contains detailed information of every DNA sequence found in human chromosomes including gene and its parts such as intron and exon. DNA sequence is tokenized following DNABERT method. There are four nucleotides: A, C, G, and T. Therefore, there are $4^3$ possible 3-mer tokens. Gene may consist of several introns and exons. This means that each of characters in sequence is labelled as either intron (i) or exon (E). Therefore, there are $2^3$ possible labels: 'iii', 'iiE', 'iEi', 'Eii', 'iEE', 'EEi', 'EiE', and 'EEE'. Intron token label and exon token label are labels given to token whose characters are all labelled 'i' or 'E', which are 'iii' and 'EEE'. Otherwise, labels are categorized into splice-site label which indicates transition from intron to exon or vice versa.

Dataset derived from DNA has high degree of imbalance because there are far less splice site tokens than exon or intron tokens. To compensate imbalance label, a weight is assigned to each label proportionate to label occurrence in data. It is also worth to note that 'EiE' and 'iEi' labels are negligible because their count in a sequence is always zero. Weight of each label is defined at (1). First definition of weight function $w$ is defined to prioritize rare labels over abundant labels. Since splice site label counts are far lower than intron and exon labels, their respective weight is almost 1. Intron and exon labels, on the other hand, are far more abundant in data and consequently their respective weight is far lower. Second definition of (1) further deprioritizes near-absent or absent labels. In this case, such labels are 'iEi' and 'EiE'. This definition introduces variable $K$ as de-prioritization magnitude. By assigning bigger value of $K$, these label weights become much lower. Data is also arranged such that splice site labels are distributed at every position. By using this arrangement, model can be trained to recognize splice site in any position.

Dataset is created from gene index. Therefore, data splitting is carried out at gene index. Primary gene index is divided into training-validation and test index with fraction 9:1. Sequences

$$w(x) = \begin{cases} \frac{\min(iiE, iEE, EEi, Eii)}{count_x}, x: \{iiE, iEE, EEi, Eii, iii, EEE\} \\ \frac{1}{\max(iii, iEE, iiE, EEE, EEi, Eii) \cdot K}, x: \{EiE, iEi\} \end{cases} \quad (1)$$

$$avg\ F1 = \frac{1}{|L|} \sum_L F1_x, L = \{iii, iiE, iEE, EEi, Eii, EEE\} \quad (2)$$

$$vr(x) = \begin{cases} [1\ 0\ 0\ 0], x = T \\ [0\ 1\ 0\ 0], x = C \\ [0\ 0\ 1\ 0], x = A \\ [0\ 0\ 0\ 1], x = G \\ [0\ 0\ 0\ 0], x = N \end{cases} \quad (3)$$

$$label2vec(x) = \begin{cases} [0\ 1], x = i \\ [1\ 0], x = E \\ [0\ 0], x = N \end{cases} \quad (4)$$

$$precision_{label} = \frac{True_{Label}}{True_{label} + False_{label}} \quad (5)$$

$$recall_{label} = \frac{True_{Label}}{True_{label} + False_{NOT\ label}} \quad (6)$$

$$F1\ score_{label} = \frac{2 \cdot precision_{label} \cdot recall_{label}}{precision_{label} + recall_{label}} \quad (7)$$

and their corresponding labels are created from each index. Training and validation sequences are generated by splitting training-validation data into two parts with fraction 8:2. Test sequences is generated directly from test index. This mechanism ensures that both training and validation data are derived from the same distribution while test data are from different distribution.

### B. Model

Proposed model, called DNABERT-SL, is composed of three parts: DNABERT, optional hidden layer(s), and classification layer with Softmax activation function. DNABERT-SL follows the same training process as BERT. Training hyperparameter observed during experiment are batch size, epoch, approach, optimizer, learning rates, and epsilons. Reference values are retrieved from both BERT and DNABERT implementation. Model architecture aspect is also tested. DNABERT-SL architectures tested are called *Base* and *Lin1*. *Base* architecture is based on BertForTokenClassification implementation found in Transformers with modifications on its BERT layer and number of token classes. *Lin1* architecture introduces additional dense layer after DNABERT layer followed with dropout to observe whether additional dense layer can improve model performance [26].

This research uses RNN model proposed to detect type and mutation index in human DNA [27] as baseline model. The report indicates that both BiLSTM and BiGRU produces similar results. One architecture produces better performance in some genes while the other works better on the other genes. Both BiLSTM and BiGRU are used as baseline to benchmark DNABERT-SL performance. The RNN model is adapted into two models: Baseline Basic and Baseline Kmer.

Baseline Basic uses one-character token instead of 3-mer token and therefore smaller vocabulary. It uses Voss representation to generate embedding in which every nucleotide is mapped into four-dimensional binary vector.

TABLE I
F1 SCORE OF DNABERT-SL, FINE TUNING, LEARNING RATE 1.10-5, EPSILON 1.10-8

| LABEL | ARCHITECTURE | | | | | | GAIN | | |
|---|---|---|---|---|---|---|---|---|---|
| | BASE | | | LIN1 | | | | | |
| | PRECISION | RECALL | F1 SCORE | PRECISION | RECALL | F1 SCORE | PRECISION | RECALL | F1 SCORE |
| iii | 0.999 | 0.996 | 0.997 | 0.999 | 0.998 | **0.998** | 0.00% | 0.20% | 0.10% |
| iiE | 0.722 | 0.999 | 0.838 | 0.797 | 0.987 | **0.882** | 10.39% | -1.20% | 5.25% |
| iEi | 0 | 0 | 0 | 0 | 0 | 0 | N.A. | N.A. | N.A. |
| Eii | 0.847 | 0.992 | 0.914 | 0.873 | 1 | **0.932** | 3.07% | 0.81% | 1.97% |
| iEE | 0.676 | 0.997 | 0.806 | 0.735 | 0.991 | **0.844** | 8.73% | -0.60% | 4.71% |
| EEi | 0.826 | 0.992 | 0.901 | 0.873 | 0.999 | **0.932** | 5.69% | 0.71% | 3.44% |
| EiE | 0 | 0 | 0 | 0 | 0 | 0 | N.A. | N.A. | N.A. |

Considering input is not always in the same size, special padding token is introduced and mapped into four-dimensional zero-valued vector. For simplicity, Voss Representation is denoted by *vr* function (2). Another aspect to be considered is label. RNN uses five label to indicate mutation type while splice site prediction requires model to recognize intron and exon. Therefore, there are two labels required, which are intron and exon. Each label is represented as two-dimensional vector as defined *label2vec* function (3).

Baselike Kmer accepts 510 3-mer tokens as input. To accommodate such input, modification is done at input layer, embedding matrix, output layer, and label vectors. Input layer is modified to accept 510 tokens. New embedding matrix is developed to uniquely represent every 3-kmer token available. There are $4^3$ possible tokens from four nucleotides. By using the same idea from Voss Representation, one-hot encoding can be applied to generate 64 unique 64-dimensional sparse-binary vectors for 64 tokens [28]. Label structure also needs to be modified. By using one-hot encoding, eight sparse-binary eight-dimensional vectors can be generated for eight token labels. To cater possible padding, another zero-valued vector is added to token and label vector collection with their respective dimension. Both Baselines and DNABERT-SL are evaluated with precision, recall, and F1 score. All evaluation metrics are computed for all labels. Precision, recall, and F1 score for each label are defined at (4), (5), and (6). Baseline Basic, Baseline Kmer, and DNABERT-SL architecture are presented at Figure 1.

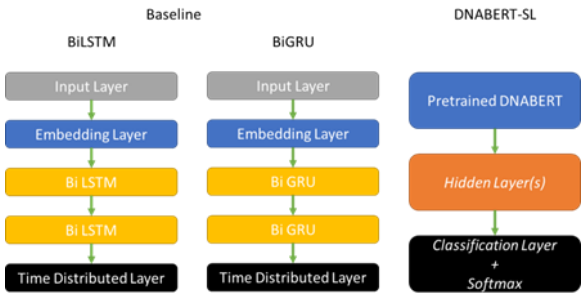

Figure 1 Baseline and DNABERT-SL Architecture

## III. RESULT

### A. Training and Validation

Experiment shows that fine-tuning produces the best DNABERT-SL model. Best training configuration for the model is as following: batch size = 32, epoch = 5, learning rate = $5.10^{-5}$, epsilon = $10^{-8}$, dropout = 0.1, and using Lin1 architecture. Experiment shows that smaller epsilon ($10^{-8}$) leads to better model as opposed to $10^{-6}$ which decreases model performance as high as 30%. Additional dense layer proves to improve model performance. F1 score on intron and exon label prediction increase 0.1% and 0.2%. F1 score on splice prediction increases 3.84% on average. DNABERT-SL with Lin1 configuration is then selected for testing (TABLE I).

TABLE II
VALIDATION AVERAGE F1 SCORE OF BASELINE BASIC AND KMER

| Epoch | Basic | | Kmer | |
|---|---|---|---|---|
| | BiLSTM | BiGRU | BiLSTM | BiGRU |
| 5 | 0.866 | 0.866 | 0.990 | 0.995 |
| 10 | 0.866 | 0.866 | 0.996 | 0.999 |
| 15 | 0.866 | 0.866 | 1 | 1 |
| 20 | 0.865 | 0.866 | 0.985 | 1 |

Validation result of Baseline Basic and Baseline Kmer can be seen at TABLE II. Baseline Basic does not seem to benefit from training as indicated by its constant average F1 score for all epochs. Unlike Baseline Basic, Baseline Kmer does learn from data as seen at decreasing loss and increasing F1 score. Both BiLSTM and BiGRU perform similarly except sudden increasing loss at the last epoch for BiLSTM. From this observation, it is safe to conclude that BiGRU performs slightly better than BiLSTM. This implies that 3-mer provides better representation for sequential labelling purpose. However, it is possible that model and training configuration mentioned at [27] does not suit the sequential labelling task for predicting splice site. Based on validation performance, Baseline Kmer is selected for testing.

### B. Test

Since Baseline Kmer performs better than Baseline Basic, Baseline Kmer BiLSTM and BiGRU are tested alongside DNABERT-SL Lin1. Comprehensive result can be seen at TABLE III. All models experience performance drop compared to their respective performance on validation data. Significant F1 score difference is observed at label 'iEE' in which BiLSTM and DNABERT-SL obtain F1 score 0.321 and 0.109 respectively, while BiGRU manages to achieve 0.575. Exon has the second most label count in data. However, all models do not seem able to predict exon successfully. BiLSTM, BiGRU, and DNABERT achieve only 0.487, 0.574, and 0.499 respectively. In case of intron prediction, all models share similar acceptable performance (0.855±0.011). In case of splice site labels, all models also share similar performance with average F1 score approximately 0.5, excluding 'iEi' and 'EiE' label.

TABLE III
Performance of DNABERT-SL and Two Baseline Models in Test Data

| LABEL | MODEL | | | | | | | | |
|---|---|---|---|---|---|---|---|---|---|
| | BiLSTM | | | BiGRU | | | DNABERT-SL | | |
| | PRECISION | RECALL | F1 SCORE | PRECISION | RECALL | F1 SCORE | PRECISION | RECALL | F1 SCORE |
| iii | 0.832 | 0.88 | 0.855 | 0.876 | 0.813 | 0.843 | 0.839 | 0.901 | **0.869** |
| iiE | 0.638 | 0.338 | 0.442 | 0.593 | 0.456 | 0.516 | 0.460 | 0.657 | **0.541** |
| iEi | 0 | 0 | 0 | 0 | 0 | 0 | 0 | 0 | 0 |
| Eii | 0.487 | 0.404 | 0.442 | 0.597 | 0.555 | 0.575 | 0.505 | 0.736 | **0.599** |
| iEE | 0.433 | 0.255 | 0.321 | 0.598 | 0.452 | **0.515** | 0.059 | 0.678 | 0.109 |
| EEi | 0.513 | 0.382 | 0.438 | 0.617 | 0.536 | 0.574 | 0.482 | 0.719 | **0.577** |
| EiE | 0 | 0 | 0 | 0 | 0 | 0 | 0 | 0 | 0 |
| EEE | 0.539 | 0.444 | 0.487 | 0.520 | 0.641 | **0.574** | 0.614 | 0.420 | 0.499 |

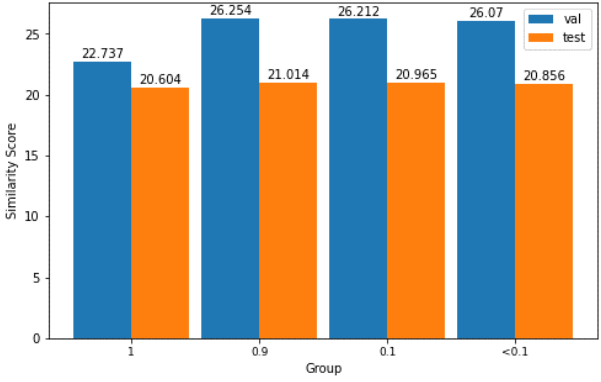

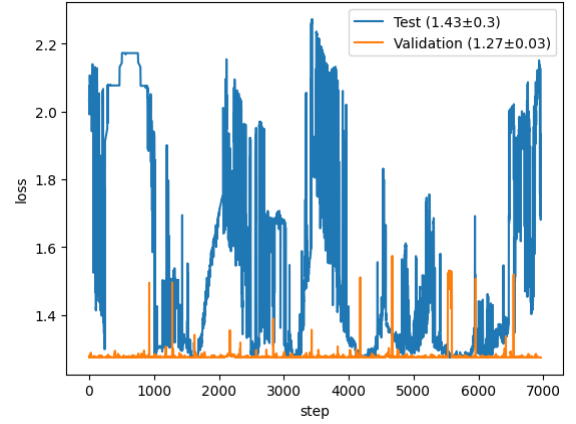

Figure 2 Similarity Scores Comparison

Figure 3 Loss Score Comparison between Test and Validation Comparison

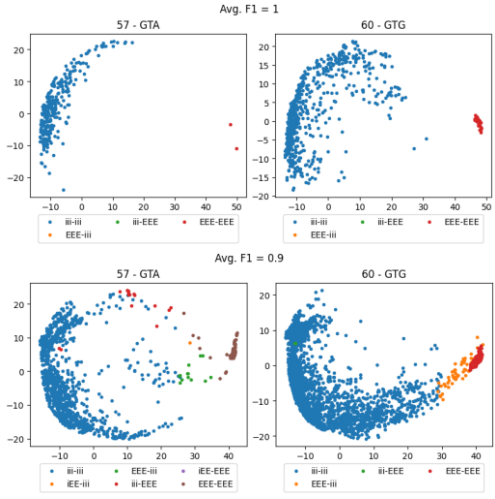

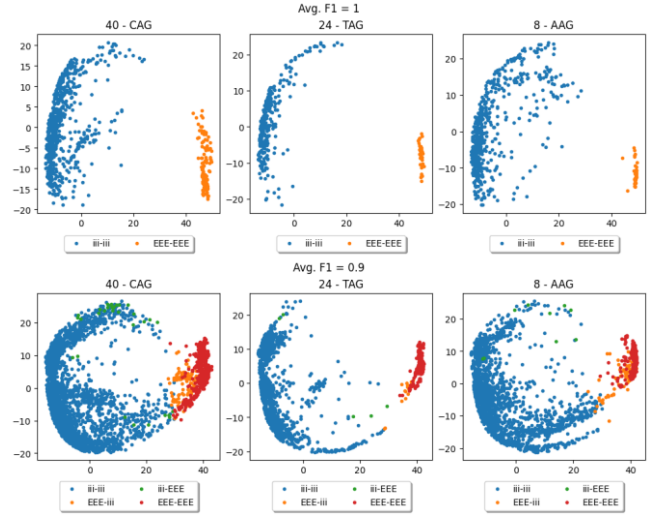

Figure 4 Motif GT Token (left) and AG Token (right)

Analysis called pairwise alignment is conducted to measure similarity between test and training data, and between validation and training data. We used Smith-Waterman [29] algorithm for pairwise alignment. It is hypothesized that model fails to recognize test data because of low similarity between training and test data. Low similarity may imply that there are key features in training data which are not found in test data. Four data groups are generated from validation and test data based on average F1 score. The four scores are 1, 0.9, 0.1, and less than 0.1. These scores are selected to represent the best and worst performing data. Forty-five sample instances are selected from each score group.

Figure 2 shows that there is small similarity score difference between all groups. The best performing validation and test data has the least similarity score difference. However, the figure also shows that validation data with F1 score less than 0.1 have higher similarity than validation data with F1 score = 1. This implies that the lesser similarity score different does not mean better prediction. On the other hand, small similarity score different intuitively infers that sequences in comparison are similar. Consequently, sequences in validation data are mostly similar with test data. Intuitively, loss scores between validation and test data should not be too different. However, DNABERT-SL shows otherwise. Figure 3 shows that validation data have significantly lower loss score than test data.

In addition to pairwise analysis, we also conducted token and motif analysis. Token analysis aims to evaluate token contextual representation. Token context is represented by two values computed from 768 values of DNABERT last layer with principal component analysis [30]. For token analysis, trinucleotides containing GT and AG are extracted from test sequences. Motif analysis is aimed to observe canon splice site motif prediction in test sequences. Motif analysis utilizes fifteen-nucleotides long sequences which have GT and AG as splice sites. These analyses are carried out at two test data group. First group has average F1 score = 1 (Avg. F1 = 1) and second group has average F1 score = 0.9 (Avg. F1 = 0.9).

Token analysis result is presented at Figure 4. We focused our analysis on canon splice site motif GT-AG. The figure presents GTA and GTG token as donor tokens, and CAG, TAG, and AAG as acceptor tokens. First group visualization shows model can distinguish exon and intron completely. In second group, model displays misclassification by presenting misclassified tokens in area between intron and exon. Similar behavior is also observed at acceptor token. Misclassified tokens can be recognized by their location in area between exon and intron area. Misclassified tokens are located near their prediction area instead of target area i.e., green-colored TAG token labelled as iii-EEE explains that TAG is labelled as exon while predicted as intron. In other words, all nucleotides in TAG are labelled as intron (i) while model predicts all nucleotides as exon (E). The same explanation also applies to GTA and GTG token. For example, Fig. 12 also shows that green-colored GTA token is labelled as EEE-iii. This indicates that all nucleotides in GTA token are labelled as intron in data while model predicts all nucleotides as exon.

Motif analysis is carried out by selecting test subsequences containing splice sites with the highest number of misclassifications for both donor and acceptor. Similarity score is computed for each of subsequences against training data with Smith-Waterman algorithm. Sequence with donor motif CAAGATCGGCCCG**GT** and sequence with acceptor motif TCTAC**AG**GTACAGAA are found to have the highest count of misclassifications. GT in donor motif is labelled as intron and therefore, all nucleotides before GT are exons. AG in acceptor motif is labelled as intron and therefore, all nucleotides before AG are intron and all nucleotides after AG are exon.

While training data provides similar motif, model still has difficulties to predict GT-AG splice site motif. In case of donor motif, model cannot distinguish GT exon and GT intron. One of many false predictions occurred to donor motif sequence is that all nucleotides in the sequence are predicted as intron. This implies that GT is not recognized as splice site motif and therefore, all nucleotides preceding GT are misclassified. To have GT as intron, all nucleotides preceding GT must be exons. Another case related to this sequence is that all nucleotides are classified as exon.

In case of acceptor motif, the correct prediction of AG is that AG is labelled as intron, all nucleotides preceding it are also intron, and all nucleotides succeeding AG are exon. However, in case of TCTAC**AG**GTACAGAA, GT after AG in acceptor motif can be incorrectly predicted as beginning of splice site, or intron. Therefore, all nucleotides after GT are intron.

Both token and motif analysis indicate that DNABERT-SL model cannot perform well in splice site motif prediction. Token analysis reveals that contextualized representation provided by fine-tuned DNABERT is not suitable to distinguish polysemy token such as GTA and GTG, or CAG, TAG, and AAG. Motif analysis reveals that prediction on nucleotides preceding splice site motif GT-AG determines success of GT-AG prediction.

IV. CONCLUSION

Splice site prediction using sequential modelling with pretrained DNABERT-3 has been presented at this paper. Experiment concludes that fine-tuning with learning rate = $5.10^{-5}$ and epsilon = $10^{-8}$ produces the best model. Additional hidden layer does increase model general performance. Test concludes that despite having high performance on validation data, model fails to achieve similar results on test data. Model often mistakenly predict splice site and exon tokens as intron token. Token analysis reveals that model cannot distinguish splice site motif based on contextual information provided by DNABERT. Motif analysis also reveals that success of GT-AG prediction depends on prediction of preceding nucleotides.

ACKNOWLEDGMENT

This paper is a representation of master thesis conducted by first author at School of Electrical Engineering and Informatics, Bandung Institute of Technology. Source code can be accessed at https://github.com/anwari32/sequence-processing.

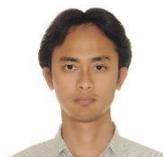

**MUHAMMAD ANWARI LEKSONO** finished his bachelor degree in informatics in October 2012 from Institut Teknologi Bandung, Indonesia. During his bachelor degree, he expressed interest in Cryptography and obtained his bachelor degree with thesis titled "Email client application with rabbit algorithm for Android smart phone" in Cryptography domain. His thesis was also published with same title. He works as System Analyst at Rekalogi, Indonesia after finishing his previous role as Policy Analyst. During his time as both Policy Analyst, he supported several government's environmental projects by contributing to system dynamics model development to analyze government public policies related to carbon emissions. He held main resposibility for Indoclimos and Syloca development as a tool for policy simulation. He is now a master graduate student at Institut Teknologi Bandung. His research interests include artificial intelligence and its applications.

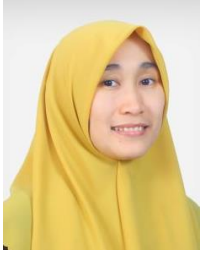# 

**AYU PURWARIANTI** was graduated from PhD program at Toyohashi University of Technology in December 2007 with dissertation title of "Cross Lingual Question Answering System (Indonesian Monolingual QA, Indonesian-English CLQA, Indonesian-Japanese CLQA)". The dissertation was in the area of Natural Language Processing or also known as Computational Linguistics which is part of Artificial Intelligence knowledge domain. Since then, she has worked as a lecturer at ITB (Bandung Institute of Technology). Other than teaching and doing research, her other activity is in Indonesian Association for Computational Linguistics where she was elected as the chair for 2016-2018; and she was also the chair of IEEE Education chapter of Indonesian section for 2017-2019. She has joined IABEE since 2015 until now. She also founded a start up named Prosa.ai since 2018. She is now the Chair of Artificial Intelligence Center at ITB since August 2019.